\documentclass[aps,pra,reprint,groupedaddress,footinbib]{revtex4-1}
\usepackage{dsfont}
\usepackage{amsmath} 
\usepackage{graphicx}
\usepackage{multirow}

\usepackage[latin1]{inputenc} % Input of ���

\begin{document}

%%%%%%%%%%%%%%%%%% title page information %%%%%%%%%%%%%%%%%%
\title{Absorption Imaging of Ultracold Atoms on Atom Chips}

\author{David~A.~Smith$^{1}$, Simon~Aigner$^{2}$, Sebastian~Hofferberth$^{2,3}$, Michael~Gring$^{1}$, Mauritz~Andersson$^{2,4}$,
  Stefan~Wildermuth$^{2}$, Peter~Kr\"uger$^{2,5}$,
  Stephan~Schneider$^{1}$, Thorsten~Schumm$^{1}$, J\"org~Schmiedmayer$^{1,2}$}

\affiliation{$^1$ Vienna Center for Quantum Science and Technology, Atominstitut, TU Wien, Vienna, Austria}
\affiliation{$^2$ Physikalisches Institut, Universit\"at Heidelberg, Heidelberg, Germany}
\affiliation{$^3$ Harvard-MIT Center for Ultracold Atoms, Harvard University, USA}
\affiliation{$^4$ School of Information and Communication Technology, KTH, Stockholm, Sweden}
\affiliation{$^5$ Midlands Ultracold Atom Research Centre, University of Nottingham, UK}

%\email{schmiedmayer@atomchip.org} %% email address is required
%\homepage{http://www.atomchip.org} %% author's URL, if desired


%\date{today}

\begin{abstract}
Imaging ultracold atomic gases close to surfaces is an important tool for the detailed analysis of experiments carried out using atom chips.
We describe the critical factors that need be considered, especially when the imaging beam is purposely reflected from the surface. In particular we present methods to measure the atom-surface distance, which is a prerequisite for magnetic field imaging and studies of atom surface-interactions.
\end{abstract}

\maketitle

%\ocis{(020.0020) Atomic and molecular physics,   
%      (110.0110) Imaging systems} % REPLACE WITH CORRECT OCIS CODES FOR YOUR ARTICLE

%%%%%%%%%%%%%%%%%%%%%%% References %%%%%%%%%%%%%%%%%%%%%%%%%
%\begin{thebibliography}{99}
%\bibliography{AtomChipImaging_final}
%\bibliographystyle{osajnl}
%\end{thebibliography}

%%%%%%%%%%%%%%%%%%%%%%%%%%  body  %%%%%%%%%%%%%%%%%%%%%%%%%%
\section{Introduction}
Microtraps created above microfabricated surfaces, or atom chips \cite{AtomChipBook,Fortagh2007,Folman2002}, are a promising approach towards the precise manipulation of ultracold atoms.  A variety of trapping, guiding and transporting potentials have been realised using current-carrying wires \cite{Sch95,Reichel1999,Mue99,Folman2000,Dekker2000,Cassettari00,hans01conveyor}, atom manipulation with electric fields  \cite{Denschlag1998,Krueger2003}, and formation of Bose-Einstein condensates (BECs) \cite{Ott2001,Haensel2001b,Leanhardt2002,Sinclair2005} have been demonstrated. Ultracold atoms held in close proximity to the chip surface are versatile probes for atom-surface interactions \cite{Lin04}, local magnetic fields \cite{Fortagh2002,Leanhardt2003,Jones2004,Esteve2004,Wildermuth2005b,Guent05} and current flow irregularities \cite{Aigner2008}. Microwave and radio-frequency (RF) fields  have been employed to coherently manipulate internal \cite{Treutlein2004} and external states \cite{Hofferberth2006} leading to interferometry with trapped atoms \cite{Schumm2005b,Wang2005,Boehi2009}, enabling routes to quantum information processing \cite{Calarco2000,Treutlein2006} and the study of 1d quantum many-body systems \cite{Hofferberth2007b,Armijo2010,VanAmerongen2008}.

Atoms are held and manipulated at short distances (a few microns) from the chip surface.  A main measurement tool is absorption imaging \cite{Ketterle1999b} and a thorough understanding of the disturbances caused by the close by atom chip surface is essential for the analysis of experiments. In this paper, we describe the key ingredients for imaging atomic clouds close to a surface and give examples of specific implementations.  We cover three different scenarios, two of which involve reflecting the imaging beam from the atom chip surface, where reflecting at grazing incidence produces a standing wave. 

\section{Basics of Absorption Imaging}\label{sec:AbsTheory}

Absorption imaging, where the attenuation of a laser beam passing through an atom cloud is measured, is the workhorse of ultracold atom experiments \cite{Ketterle1999b}.  The shadow cast by the atom cloud onto the CCD allows an estimation of the atomic column density. In the case of low atom numbers, the laser light is chosen to be resonant with an atomic transition, whereas off-resonant phase-contrast imaging offers a non-destructive imaging alternative for large column densities (large atom numbers).  Here we only consider imaging with resonant light.

\subsection{Basic Optical Setups}
\begin{figure*}
  \centering \includegraphics[width=17cm]{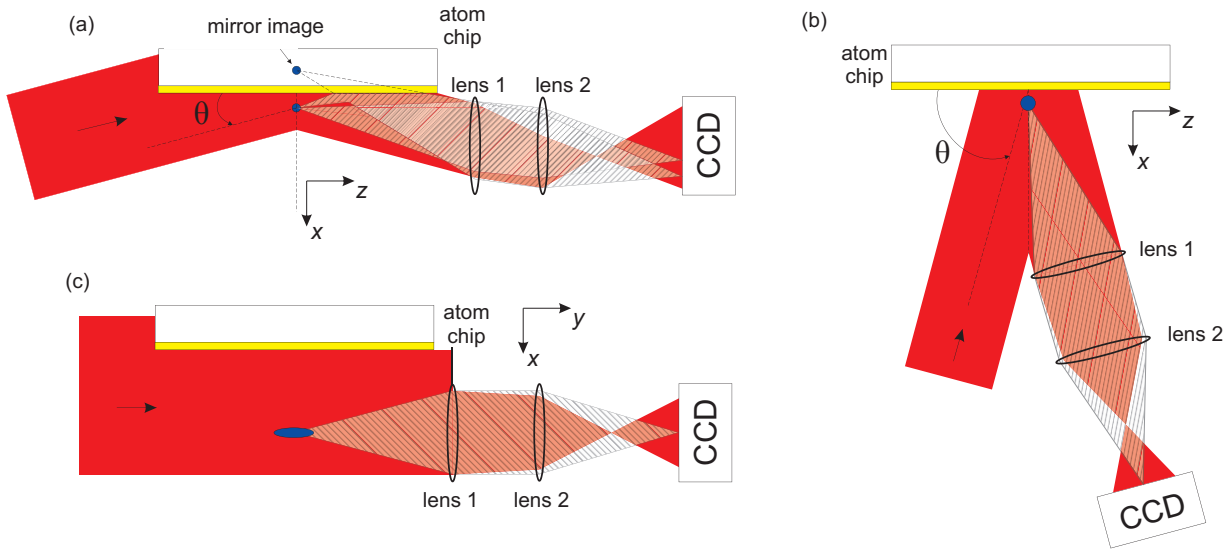}
  \caption{Sketch of Imaging Configurations (not to scale). (a) \textit{Grazing Incidence Imaging}. The imaging beam is reflected off the atom chip surface and the atom cloud (blue circle) produces two shadows. For the direct image, the imaging beam first reflects off the atom chip surface, then interacts with the atom cloud.  For the mirror image, the imaging beam first interacts with the atom cloud, then reflects off the atom chip. (b) \textit{Normal Incidence Imaging} provides the location of the atom cloud relative to the structures on the atom chip surface. (c) \textit{Time-of-flight Imaging}, where the atom cloud is far from the atom chip surface.}
  \label{fig:translong}
\end{figure*}

Figure~\ref{fig:translong} shows a schematic of three different imaging configurations. When imaging atoms close to the chip surface \cite{PhysRevA.67.023612} (Figs.~\ref{fig:translong}(a) and~\ref{fig:translong}(b)) diffraction and/or reflection of the imaging beam is unavoidable . For images taken far from the chip (Fig.~\ref{fig:translong}(c)) the influence of the surface becomes negligible and this configuration operates as in standard absorption imaging. In the rest of the paper $\theta$ describes the angle of incidence of the imaging beam with respect to the chip surface.  For $\theta=0^\circ$ the imaging beam propagates parallel to the surface of the atom chip. For $0 < \theta \leq 90^\circ$ the imaging beam is reflected from the surface.

A typical imaging configuration uses a two-lens setup to focus the image of the atoms onto the CCD. Lens 1 is positioned at (or close to) its infinity optical working point from the atom cloud (such that the lens collimates the diffracted light) and lens 2 is used to form the image.

The atomic samples on atom chips are usually in a very elongated quasi-one-dimensional configuration. In \textit{longitudinal} imaging the probe beam is parallel to the elongated extent of the atom cloud and integrates over the full length of the cloud.  In \textit{transversal} imaging the probe beam is orthogonal to the long axis of the cloud, allowing the study of properties along the quasi-1d cloud.  The short extent of the cloud in the transversal direction, especially for in situ images or after a short time of flight, allows a relatively small depth of field for transversal imaging, and consequently higher resolution.  In contrast, the resolution of a longitudinal imaging system is limited by the large extent of the cloud which requires a depth of field up to several hundred microns, thus putting a lower bound on the achievable resolution.
We define here the axes convention that will be used below: the $x$-direction is the direction of gravity, the $y$-direction is along the longitudinal direction of the atom cloud, and the $z$-direction is along the (horizontal) transverse direction of the atom cloud.

\subsection{Resonant Atom-Light Interaction}
In our absorption imaging we use resonant light. If the incoming intensity $I_\mathrm{in}$ is sufficiently below saturation, the attenuation of the incident probe beam is given by
\begin{equation}
    \frac{I_\mathrm{out}}{I_\mathrm{in}}=\exp(- n \sigma).
    \label{eq:absorption}
\end{equation}
$I_\mathrm{out}$ is the outgoing (i.e. unscattered) intensity, $n$ the column density of the atoms and $\sigma$ the absorption cross section of the specific atomic transition. For large intensities $\sigma$ has to be multiplied by a factor $1/(1+I_\mathrm{in}/I_\mathrm{sat})$ to account for saturation effects ($I_\mathrm{sat}$ is the saturation intensity of the optical transition).  Estimating $n$ relies on knowing both $I_\mathrm{in}$ and $I_\mathrm{out}$.  Experimentally, a CCD camera is used to take two pictures: one without atoms giving $I_\mathrm{in}$ and one with atoms giving $I_\mathrm{out}$, from which one can then calculate the density profile of the atom cloud using eq.~\ref{eq:absorption}.  The division of the two images ($I_\mathrm{out}/I_\mathrm{in}$) mostly eliminates the effects of any inhomogeneous spatial intensity distribution. However, ultimately, detection is limited by photon shot noise~\cite{Bucker2009}.

\section{Experimental Implementation}\label{sec:ChipSetup}
The imaging systems described here have been implemented in atom chip setups in Heidelberg/Vienna. We typically load $> 10^6$ cold ($< 10 \mu \mathrm{K}$) Rb atoms into a selected chip trap. Using radio-frequency evaporative cooling we create either an ultracold thermal cloud or a BEC.

Our atom chips have been discussed in detail in~\cite{Groth2004,Trinker2008}.  For imaging it is important that their high-quality gold surfaces are exceptionally clean and excellent mirrors.  Scattered light comes predominantly from the thin etchings (typically $<\,10 \, \mu$m) that define the wire structures. Multilayer atom chips \cite{Trinker2008} may have distinct surface profiles that have to be taken into account.

The imaging light is guided to the experiment using a single-mode polarization-preserving optical fibre.  The light expanding from the fibre core is collimated using a precision achromat and sent through a high-quality optical window into the vacuum chamber containing the atom chip and the atom cloud. In most of our imaging systems we use optics built from two lenses, as in Fig. \ref{fig:translong}. For details on specific systems, see section~\ref{sec:TwoLens}.

\begin{figure*}[t]
  \centering {\includegraphics[width=16cm]{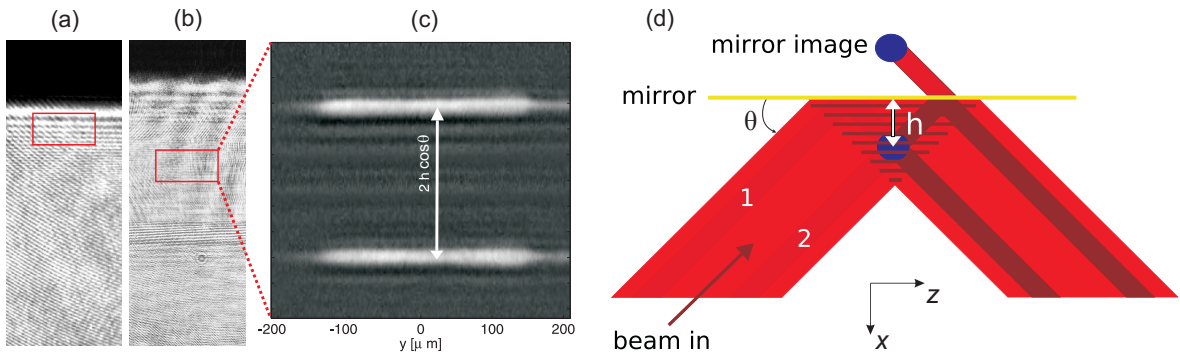}}
  \caption{(a,\,b) Images without atoms ($I_\mathrm{in}$) with (a) the imaging beam propagating parallel to chip surface and (b) the imaging beam reflected from chip surface with $\theta = 2^{\circ}$. The dark region at the top is the shadow cast by the chip and its mounting.  The red boxes represent the region where the atom cloud would be located in situ or for short times of flight.  The strong horizontal fringes are due to diffraction from the edges of the chip. The inclination of the beam allows the diffraction effects to be moved in the image relative to the position of the atom cloud. The circular fringes and other structures are due to small dust particles in the beam path.
(c,\,d) Reflecting the imaging beam off the atom chip surface results in a standing wave (d) and 2 clouds in the image (c) because the atom cloud (blue circle) is passed by two different beam paths. Path (1) is mapped by the imaging system to a real image, path (2) to a mirror image.  The image in (c) would lie in the area denoted in (b).}
  \label{fig:ShadowsIm}
  \label{fig:Fig5p2}
\end{figure*}

\section{Grazing-Incidence Imaging} \label{sec:TransIm}
The propagation of the imaging beam (almost) parallel to the atom chip surface results in diffraction from the surface edge, as seen in Fig.~\ref{fig:ShadowsIm}(a).  Even though intensity variations are vastly reduced in the final absorption image, residual noise much larger than photon shot noise usually remains.  A more uniform image in the region of the atom cloud can be achieved by inclining the imaging beam such that it reflects from the chip surface (Fig.~\ref{fig:translong}(a))~\cite{PhysRevA.67.023612}, moving the diffraction effects within the image, as can be seen in Figs.~\ref{fig:ShadowsIm}(a) and~\ref{fig:ShadowsIm}(b).

When the imaging beam is reflected from the chip surface, two beam paths traverse the cloud (Fig.~\ref{fig:ShadowsIm}(d)).  One first passes through the atom cloud, then reflects from the atom chip surface.  The second path first reflects from the atom chip surface, then passes through the atom cloud. This creates two images, their separation $d$ measures the distance $h$ of the atoms from the chip surface ($d=2h\cos\theta$). At an angle of incidence of $\theta \sim 2^\circ$, one can typically observe two-shadow images for cloud-surface distances up to $\sim300 \mu$m.  At very small $h$ the two shadows merge into one, dictated by the resolution of the imaging system.

At grazing angles of incidence one has to consider carefully the polarization of the imaging beam. Reflection from the metallic gold surface of the chip produces different phase shifts for the components of the electric field oscillating in-plane and out-of-plane of the surface and hence the polarization is in general not preserved. Moreover, the phase shift of the out-of-plane component of the electric field depends strongly on the angle of incidence to the surface. Therefore an incident imaging beam with general polarization can lead to a poorly defined imaging situation in which both the intensity and polarization change with distance to the surface.

To simplify the understanding of the imaging process, we focus on the cases of linearly polarized light oscillating in-plane or out-of-plane of the chip surface. In both situations the polarization is preserved but one has to deal with a standing light wave above the surface (Fig.\,\ref{fig:Fig5p2}). Consequently, a simple geometrical picture of imaging by reflecting the imaging beam from the surface neglects important details which need to be considered in detail.

\subsection{Reflection and Standing Waves} \label{sec:Reflec}

The illumination of the atomic cloud by a standing wave $I_{\mathrm{sw}}(x)=4 I_{\mathrm{in}}\sin^2(k_x x)$ where $I_{\mathrm{in}}$ is the intensity of a single beam and $k_x= k \sin{\theta}$ is the wavevector component perpendicular to the surface, is highly inhomogeneous and in stark contrast to standard absorption imaging \cite{Ketterle1999b}. The effect of this inhomogeneous illumination depends on the atom column density profile $n(x)$.  The local scattered intensity $I_\mathrm{sc}(x)$ is given by:
\begin{equation}
    I_\mathrm{sc}(x)=I_{\mathrm{sw}}(x) \left[1- e^{ - n(x) \sigma_\mathrm{sc}} \right].
\end{equation}
The total scattered power is then given by $P_{\mathrm{sc}}=\int I_\mathrm{sc}(x) dx $. If we assume an atom cloud with a gaussian density profile $n(x)=n(h) \exp\left[-(x-h)^2/(2w^2)\right]$ of width $w$ located at height $h>>w$ and weak absorption ($n(h) \sigma_\mathrm{sc}<<1$), then the integral can be solved analytically and we obtain
\begin{equation}
    P_{\mathrm{sc}} = 2 P_{0} [1 -  \cos{(2 k_x h)} e^{- 2 k_x^2 w^2}],
     \label{eq:ScattPow}
\end{equation}
where $P_0$ is the power that would be scattered by the same absorber in a single uniform plane wave of intensity $I_{\mathrm{in}}$. As the cloud is moved through the standing wave, it probes the local intensity. Depending on the size of the cloud, it truly samples the intensity, or averages over a broader range. 

For $k_x w~<<~1$ (i.e., the cloud is small compared to the period of the standing wave), the scattered power is determined by the local intensity at the position of the atoms and is therefore modulated depending on the position within the standing wave. As the phase of the standing wave is fixed by the surface, this modulation of the scattered power can be used as a reference ruler for measuring the distance to the surface (see section~\ref{sec:Ruler}).

For $k_x w~>>~1$, the scattered power becomes independent of the position within one period of the standing wave.  It approaches twice the amount obtained from illumination by a single plane wave of intensity $I_{\mathrm{in}}$. This is the regime of geometric optics. The light passing through the cloud can be thought of as coming from two beams (Fig.~\ref{fig:Fig5p2}(e)), where one beam hits the cloud directly and the other one is first reflected by the chip surface. Consequently, the influence of the standing wave can be reduced by inclining the probe beam by a large enough angle so that the typical standing wave periodicity is much smaller then the transverse sizes under interest, as used in the microscopy scans in \cite{Aigner2008}.

\begin{figure*}[tb]
  \centering \includegraphics[width=11cm]{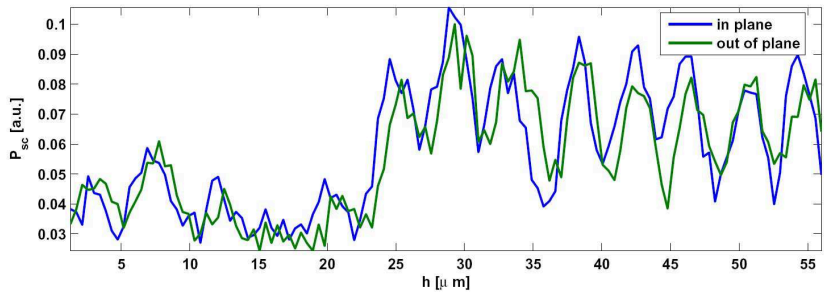}
  \caption{The total scattered power $P_{sc}$ for trapped condensates at different heights above a wire. The blue and green curves show the result for in-plane and out-of-plane linear polarization of the imaging beam, which have a small relative phase shift caused by the different boundary conditions of the standing wave at the mirror surface. The angle of incidence of the imaging beam $\theta$ was 4.2$^\circ$.  The jump in signal between $h=20$ and $h=25$ is due to wires of different heights obscuring part of the beams (see Fig.~\ref{fig:diffheights}).  Imaging System 2 in section~\ref{sec:TwoLens} was used in this case.}
  \label{fig:Fig5p3}
\end{figure*}

\subsection{The Standing Wave as a Ruler} \label{sec:Ruler}
\begin{figure*}[tb]
  \centering {\includegraphics[width=15cm]{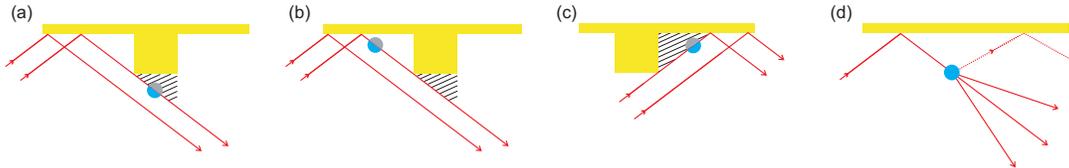}}
  \caption{(a-c). Imaging close to wires of different heights. The three scenarios show the atom cloud (blue circle) above different parts of the chip in the situation where the atom chip surface has wires of different heights.  Shadows cast by the wires into the imaging beam result in part of the cloud not being imaged in each case.
  (d) Angular aliasing. If a plane wave component scattered by the atom cloud is reflected by the surface (dashed line), it exits under an angle that is already occupied by a wave component that travels directly away from the surface.}
  \label{fig:diffheights}
  \label{fig:Fig5p6}
\end{figure*}

By varying the position of a thin cloud ($k_x w~<<~1$) and counting the minima and maxima of the detected scattered power, it should be possible to determine the absolute position above the atom chip surface.  Figure \ref{fig:Fig5p3} shows the result of such a measurement where the total scattered power $P_{sc}$ has been measured for trapped condensates at different distances from the surface. The blue curve shows the result for an imaging beam which is linearly \textit{p}-polarized in the plane of incidence, whereas the green curve is for linear \textsl{s}-polarization. As can be seen, the measurements do not follow a simple harmonic modulation: phase shifts and changes in the modulation amplitude appear. These artifacts are caused by the details of the wire structures on the chip surface. The dominant effect is the different heights of the wires (Fig.~\ref{fig:diffheights}) which, besides casting shadows, results in a phase shift of the reflected beam that is proportional to the local surface structure height. As the reflected beam has to propagate a significant distance to reach the position of the atomic cloud, diffraction has a significant influence on the observed patterns.

A second effect is connected to the non-ideal boundary conditions of a real gold surface. The phase shift that occurs upon reflection is not exactly the one expected from an ideal conductor and depends upon the angle of incidence $\theta$ and polarization (Fig.~\ref{fig:Fig5p3}).

\subsection{Angular Aliasing} \label{sec:AngAliasing}
A different modification of the image results from diffraction and scattering from the atomic cloud. Figure~\ref{fig:Fig5p6}(d) sketches the basic scenario for the situation where the incident plane wave is reflected by the mirror surface and then passes through the atomic cloud. If the size of this cloud is large, then the image is dominated by a geometric shadow that propagates along the same direction as the plane wave. However, as the cloud becomes smaller, important components of the scattered wave occupy a larger angular spread relative to the direction of the plane wave. As long as this spread is small compared to the angle between the imaging beam and the atom chip surface, all the light propagates away from the mirror: a lens will reconstruct the true wavefront in the object plane. This situation changes when the angle spread of the scattered wave becomes large enough that a significant fraction propagates towards the mirror surface. This part of the wave is reflected and wrongly mapped to an angle under which another part of the wave travels directly away from the mirror. This aliasing effect is most important in the regime of grazing incidence imaging.

\subsection{Wavefront Propagation}
The fine details of the imaging depend on the light propagation close to the surface of the chip. For a full understanding we implemented a numerical model of the absorption imaging setup, including the specific properties of the surface and the imaging lenses.  Such a model includes diffraction and scattering of the propagating wave by the chip and, of course, by the atoms.

We consider an linearly s-polarized optical plane wave propagating mainly along the $z$-direction with $k$-vector $\vec{k}=(k_x,k_y,k_z)$ and wavelength $\lambda = 2 \pi/k$. We assume that in the plane $z=z_0$ the optical amplitude is given by $U(x,y,z_0)$ in the scalar approximation. The propagator for relating $U(x,y,z_0)$ to the amplitude in a later plane $U(x,y,z_0+\Delta z)$ is given by \cite{SalehTeich}
\begin{eqnarray}\label{propagator}
    K(\nu_x,\nu_y) &= & \exp \left( -i 2 \pi \sqrt{1/\lambda^2 -\nu_x^2-\nu_y^2} \Delta z \right)
\end{eqnarray}
where $\nu_i=\sin(\phi_i)/\lambda \approx \phi_i/\lambda$ in the paraxial approximation and $\phi_i$ is the wave vector angle. Since this propagator is diagonal in the angular representation it can be conveniently implemented in the Fourier representation \cite{SalehTeich}:
\begin{widetext}
\begin{eqnarray}\label{fft}
   V(\nu_x,\nu_y,z_0)  &=&\int_{-\infty}^{\infty} \int_{-\infty}^{\infty}
         U(x,y,z_0) e^{ -i 2 \pi (\nu_x x+\nu_y y )} dx dy\\
   U(x,y,z_0+\Delta z) &=& \int_{-\infty}^{\infty} \int_{-\infty}^{\infty}
         K(\nu_x,\nu_y) V(\nu_x,\nu_y,z_0)
   e^{ -i 2 \pi (\nu_x x+\nu_y y ) } dx dy
\end{eqnarray}
\end{widetext}
i.e. $V(\nu_x,\nu_y,z_0)$ is the Fourier transform of $U(x,y,z_0)$, and the amplitude $U(x,y,z_0+\Delta z)$ at a point $z_0+\Delta z$ is related to $U(x,y,z_0)$ through the propagator $K(\nu_x,\nu_y)$. Numerically, the above can be efficiently implemented using the fast Fourier transform (FFT), which we utilise in the following calculations.

The optical propagation now consists of the following steps. The initial wavefront is considered to be a wide Gaussian (wide with respect to the chip structures) with a small adjustable angle $\theta$ to the chip surface.  The wavefront is truncated by the chip edge and the first propagation step is to the chip centre.  For propagation over the chip (mirror) surface, a reflecting boundary condition at the mirror surface is implemented with a phase shift $\delta=-\pi$ (for s-polarized light).  At the centre of the chip, the atoms absorb, at their location, part of the light. The such modified light is propagated to the end of the chip and then to the first lens in the imaging system.  The wavefront is truncated by the first lens (which leads to a finite optical resolution) and subsequently propagated through the Fourier plane, the second lens, then finally to the detector plane.  This last step also involves an adjustable defocus.  Finally, the intensity at the detector is down-sampled to the pixel resolution of the CCD.
In order to calculate the absorption signal, two propagations are made: one with atoms present (resulting in $I_\mathrm{out}$ in terms of CCD images discussed previously) and one without atoms present ($I_\mathrm{in}$).  The resulting signal is calculated in the same way as for the real experimental images.

\begin{figure*}[t]
  \centering   \includegraphics[width=13cm]{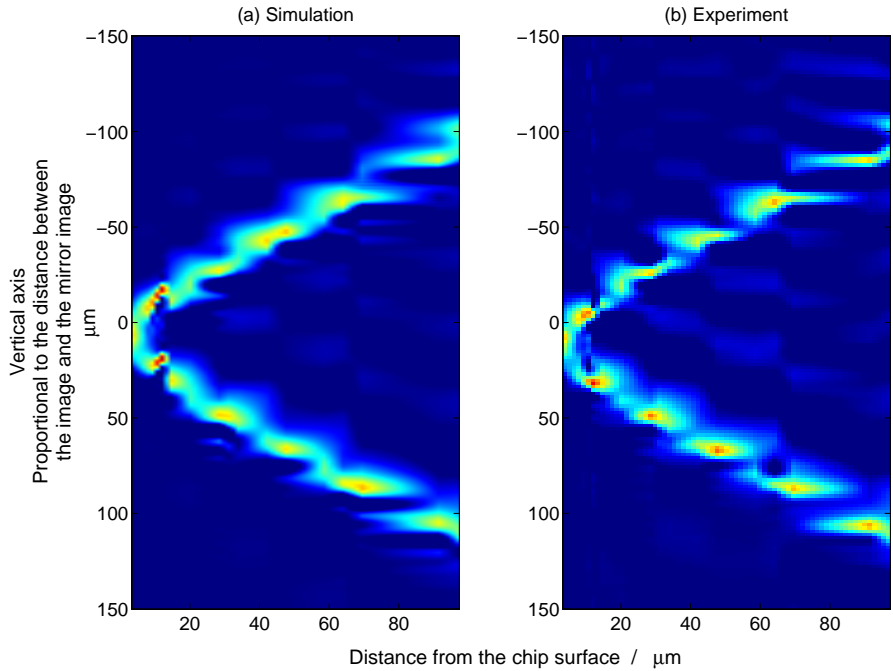}
  \caption{
(a) Simulation and (b) experimental data of absorption images for varying distance between the atom cloud and the chip surface. Shown are vertical line densities for different trap distances, with two separate absorption positions emerging as the distance between the atom cloud and the chip increases (Fig.~\ref{fig:Fig5p2}). It can be seen that many of the features resulting from interference effects due to the reflecting surface, such as dark fringes where the atoms are located, are well reproduced by the simulation. The good agreement between theory and experiment allows for an accurate calibration of the trap distance from the surface, which in turn allows for precise calibration of the magnetic fields applied to form the magnetic trap.
Imaging System 1 in section~\ref{sec:TwoLens} was used in this case.}
  \label{fig:propagation}
\end{figure*}

Figure~\ref{fig:propagation}(a) shows the results of simulating the imaging of an atom cloud at various heights above the chip surface for an actual experimental implementation. The corresponding experimental data for in situ absorption images of the atom cloud is shown in Fig.~\ref{fig:propagation}(b), where for each individual trap height the image was integrated in the direction parallel to the chip surface to obtain the line density. Experiment and simulation are in remarkable agreement. In accordance with the simple geometric picture (Fig.~\ref{fig:translong}(a)), two lines of images emerge with increasing height above the chip.  Due to the effect of the standing wave above the surface, atoms are detected with different clarity at different heights.  In addition, intricate details emerge which are related to scattering and diffraction of the imaging light from the chip and the atoms. Clearly, to find a good position for imaging, a full understanding of the wave propagation close to the surface is essential.  Only then will a comparison between experimental data and simulation allow an accurate determination of the trap height $h$ for each image.  With the precisely known current in the chip wire, this yields a very good calibration of the homogeneous magnetic bias field applied to form the atom chip traps, and faithful calculation of atom positions for other trapping parameters.

\begin{figure*}[t]
  \centering \includegraphics[width=13cm]{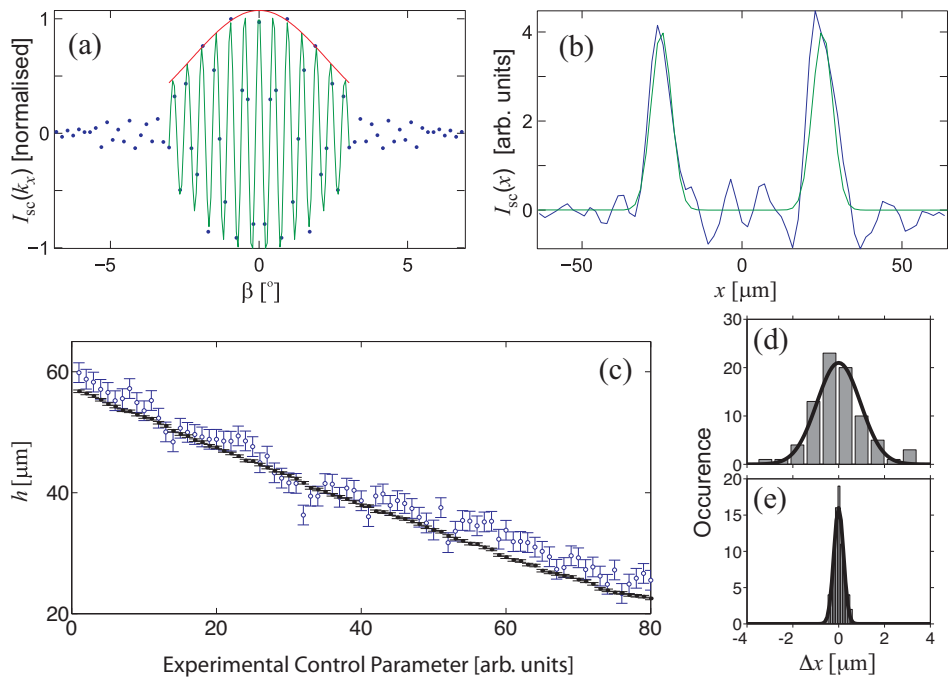}
  \caption{Extracting the height above the chip by Fourier method.
  (a) Fourier transform $\tilde{I}_\mathrm{sc} (k_x)$ of (b) the scattered intensity $I_\mathrm{sc}(x)$. The transverse wave vector $k_x$ has been translated to the propagation angle $\beta$ of the corresponding plane wave component. The green curve shows a fit of the model eq.~\ref{eq:Fourier} to the data, the red curve shows the envelope $exp(-k^2_x w^2/2)$, where $w$ is from a fit to the experimental data $I_\mathrm{sc}(x)$ shown in (b), which shows the scattered intensity profile $I_\mathrm{sc}(x)$ together with the profile obtained from the fit in Fourier space (green line).
  (c) Comparing height estimation methods: Taking the distance directly from $I_\mathrm{sc} (x)$ to obtain $h$ (open circles) and the corresponding result of the Fourier method (filled circles). The experimental control parameter is nearly linear in height above the surface and is related to the magnetic field controlling the magnetic trap. The Fourier approach presents the more stable and less noisy method.
  (d) Residuals from the direct fitting method of (b), giving a standard deviation of $0.95\,\mu$m.
  (e) Residuals from the Fourier fitting method of (a), giving a standard deviation of $0.21\,\mu$m.
   Imaging System 2 in section~\ref{sec:TwoLens} was used in this case.}
  \label{fig:Comparison}
  \label{fig:Fourier}
\end{figure*}

\subsection{Fourier Analysis} \label{sec:Fourier}
Although it is instructive to analyze the full wave propagation, Fourier analysis provides a powerful method to ascertain the distance between the atom cloud and the surface.  An experimentally determined absorption profile of the scattered intensity $I_\mathrm{sc}(x)$ can be decomposed into its Fourier components $\tilde{I}_\mathrm{sc}(k_x)$. If the measured profile is well described by two Gaussian curves of width $w$, separated by the distance $2 h$, the real part of the Fourier transform of $I_\mathrm{sc} (x)$ can then be written as follows:
\begin{equation}
    \tilde{I}_\mathrm{sc} ^\mathrm{~model} (k_x) \propto \cos{(k_x h)}~\exp{(-k_x^2w^2/2)}.
    \label{eq:Fourier}
\end{equation}
This simple model can be fitted to the Fourier transform of the experimental profile $\tilde{I}_\mathrm{sc} ^\mathrm{~exp} $. The distance $h$ is then extracted, using only the small $k_x$ component of the experimental spectrum, which is affected little by the aliasing of angles and the high frequency noise in $I_\mathrm{sc} ^\mathrm{~exp} $. This Fourier transform method can be seen as filtering the relevant information out of the images.  It extracts the information about the distance and suppresses frequency components associated with the fringes from the standing wave and other spurious interference effects.

Figure~\ref{fig:Fourier} shows a typical example of a measured profile $I_\mathrm{sc}(x)$ and its Fourier transform $\tilde{I}_\mathrm{sc}(k_x)$. In Fig.~\ref{fig:Fourier}(b) compares the original $I_\mathrm{sc} (x)$ and that calculated from the fit to $\tilde{I}_\mathrm{sc} (k_x)$. This Fourier method suppresses the high frequency noise in $I_\mathrm{sc} (x)$. Compared to a direct evaluation of $h$ from $I_\mathrm{sc} (x)$, the Fourier method is significantly more stable, as illustrated in Fig.~\ref{fig:Comparison}(c-e).  Fig.~\ref{fig:Comparison}(c) shows, for both methods, the extracted height $h$ plotted against a high-resolution, stable experimental control parameter (a precise magnetic field in this case) that is nearly linear in actual height. The error for the Fourier method (for each $h$) is smaller than that for the direct method, and the smoothness of the data for the Fourier method is in line with the behaviour expected when using the near-linear control parameter. The residuals in figs.~\ref{fig:Comparison}(d,\,e) are for the deviation of the extracted $h$ relative to this expected near-linear behaviour.  This result is a direct consequence of the removal of noise from the fitting using the Fourier method.

\section{Orthogonal-Angle-of-Incidence Imaging} \label{sec:SmallAngle}
Orthogonal-angle-of-incidence imaging ($\theta \sim 90^\circ$) is primarily used to locate the atom cloud relative to the chip structures (Fig.~\ref{fig:translong}(b)).  The imaging beam is reflected from the gold surface of the chip and hence interacts with the atom cloud twice, giving two images. For atoms close to the surface, both images are in focus and overlap with each other nearly perfectly.  However, when the atom cloud is further from the chip, the two shadows separate and in many cases only one shadow can be in focus.  With decreasing $\theta$ the two shadows also separate further laterally.

\begin{figure*}[t]
    \centering{\includegraphics[width=12cm]{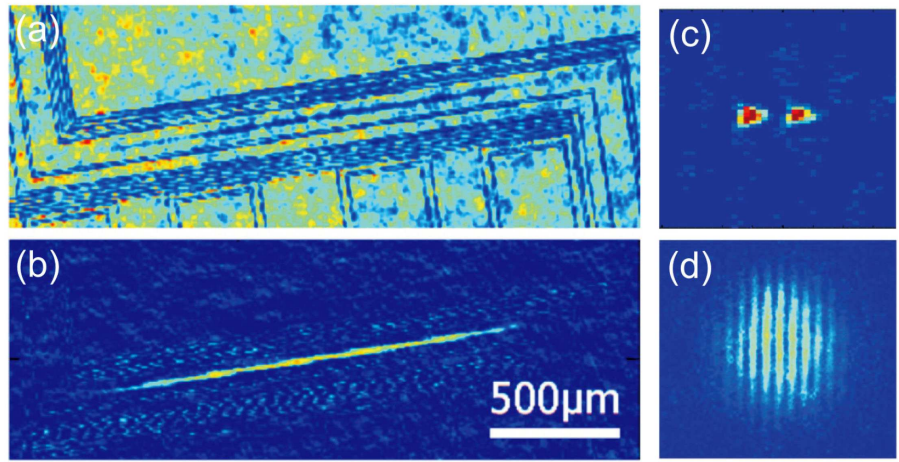}}
    \caption{\textit{(left)} Orthogonal-angle-of-incidence imaging of an atomic cloud above a broad 100 $\mu$m wide Z-shaped trapping wire.
    (a) The direct image reveals the features on the chip. The atom cloud is just visible in the center of the central broad wire.
    (b) Processed absorption picture (divided by a reference image without atoms). The atoms are clearly visible and the speckle patterns are reduced.  (Imaging System 3 section~\ref{sec:TwoLens}).
    \textit{(right)}: Longitudinal imaging. (c) in situ image, 80 $\mu$m away from the chip surface showing a BEC that has been split by $\sim$ 45 $\mu$m using a RF dressed-state double-well potential.
    (d) Image of time-of-flight matter-wave interference of two BECs after 15~ms time of flight \cite{Schumm2005b}.  (Imaging System 4  section~\ref{sec:TwoLens}).}
    \label{fig:VertImage}
    \label{fig:LongituImage}
\end{figure*}

This imaging method relies even more on high-quality reflection from the chip surface. Image quality is very sensitive to light scattering and diffraction due to residual surface roughness and the gaps between the structures defining the wires on the atom chip. The image quality also depends on the position of the atoms. It can be excellent if the atoms are far away from wire edges, for example, above the centre of a broad wire, as shown in Fig. \ref{fig:VertImage}. Images degrade for atoms located close to the wire edges because near-field diffraction effects make it difficult to interpret the images.

In our implementation of orthogonal-angle-of-incidence imaging (Fig.~\ref{fig:VertImage}), we introduce a small deviation from normal incidence ($\theta \sim 84^\circ$), just large enough so that the incoming imaging beam can pass the imaging objective lens.  The input imaging beam is sent through an optical fibre (from which it is collimated to have a diameter of approximately 2 cm) and reflected from the atom chip surface. The atom cloud absorption is imaged by a two-lens system that has a working distance of approximately 13 cm (Imaging System 3 in section~\ref{sec:TwoLens}).

\section{Imaging Far from the Chip Surface and Longitudinal Imaging} \label{sec:LongIm}
When imaging far from the atom chip surface, the imaging beam can be passed parallel to the chip surface, as shown in Fig.~\ref{fig:translong}(c).  At distances $>50$ $\mu$m from the surface, the effects of the diffraction from the chip edge (Fig.~\ref{fig:ShadowsIm}(a),(b)) and chip structures are less deleterious to the image quality.  This imaging configuration can be used for both transverse and longitudinal imaging.

The longitudinal imaging system is typically chosen to work at a lower resolution, smaller f-number, compared to the transverse imaging direction, mainly because the atom clouds have a large extension along the imaging beam, and therefore require imaging with a larger depth of field.  Additionally, in our setups, the horizontal MOT beams for initial atom cooling and the longitudinal imaging beam are overlapped with polarizing beam-splitter cubes, which places limitations on how close to the atom cloud the first imaging lens can be placed.

Figure~\ref{fig:LongituImage} shows two examples of longitudinal images. Figure~\ref{fig:LongituImage}(c) is an in situ image of a Bose-Einstein condensate (BEC) that has been split in a double-well potential 80 $\mu$m away from the chip. Figure~\ref{fig:LongituImage}(d) shows a time-of-flight image of matter-wave interference between two BECs released from a double-well potential \cite{Schumm2005b}.  In this case, the resolution of the imaging system must be good enough to observe the interference fringes whilst having a sufficiently large depth of field to keep the entire length of the cloud in focus. Imaging system 4, as detailed in section~\ref{sec:TwoLens}, was used for these images.

\section{Image Quality Optimisation}\label{sec:ImQual}
Generally, the raw CCD images contain large amplitude fringes and speckle arising from the interference of spuriously scattered light with the imaging beams~(Fig.\,\ref{fig:ShadowsIm}).  In principle, these fringes can be divided out when the attenuation is calculated from the two recorded images ($I_\mathrm{out}$ and $I_\mathrm{in}$). However, even small shifts in the position of structures in the imaging beam path (e.g. due to vibrations) can lead to significant changes in the speckle pattern and large disturbances in the resulting absorption image. To minimize these disturbances we usually implement the following:
\begin{itemize}
    \item Low-noise (in terms of loudness) light beam shutters are used to block laser light. The shutters are mechanically isolated from the laser table.
    \item All optical elements are attached as rigidly as possible onto the laser table to avoid vibrations and all stray light at optical elements in the beam path is minimised. Covering the complete imaging beam path with dust-free tubes (whether plastic or card) not only prevents optical elements from being soiled with dust but additionally avoids air turbulence.
    \item Any fans operating in the CCD cameras for cooling are operated in a pulsed mode and are switched off several seconds before the images are taken.
    \item Good overlap of the two images (atom cloud and reference images) can be reached if the delay time between the two images is reduced as much as possible.  A frame-transfer camera enables a shift of the first image within a few ms across to a masked region on the CCD chip before the second image is taken. After both images have been taken, the low-noise readout of the CCD chip then usually takes several seconds.
\end{itemize}

In addition, interference structures can originate from the interference of light being reflected between the CCD chip and vacuum window of the CCD camera. These fringes can be avoided if the CCD vacuum window is anti-reflection coated for the specific wavelength or if the camera is tilted with respect to the incoming light, however the tilting limits the field of view.  It is also possible to use advanced post-processing techniques to remove fringes in images~\cite{Ockeloen2010}.

\begin{figure*}[t]
  \centering \includegraphics[width=12cm]{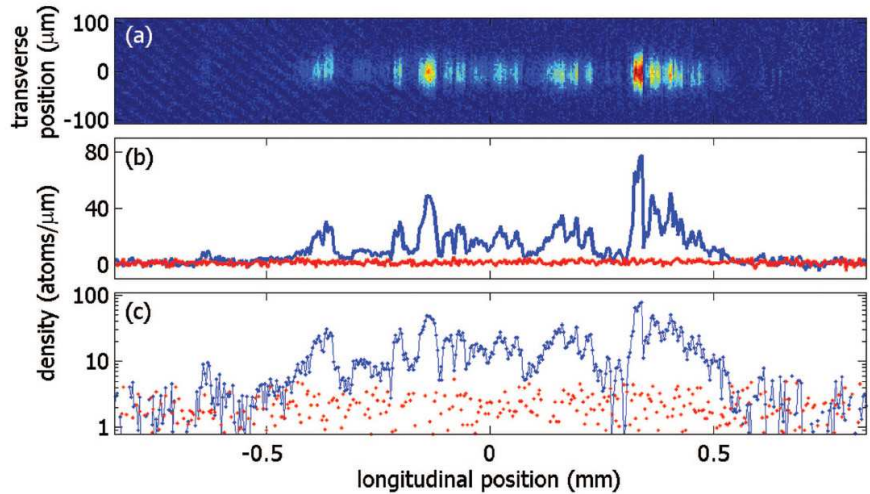}
  \caption{(a) Absorption image of a fragmented BEC taken after 5 ms of time-of-flight expansion. The BEC has been formed at a distance of 10 $\mu$m from the wire surface. (b) Longitudinal one-dimensional density profile (blue) derived from the absorption image. The noise-level is shown in red. (c) To accentuate the noise-floor of $\sim$ 2~atoms/$\mu$m rms, the same data has been plotted logarithmically. Imaging System 1 in section~\ref{sec:TwoLens} was used for this image.}
  \label{fig:noiseTOF}
\end{figure*}

The noise level in our images can be estimated from a region where no atoms are present.  In images far from the chip surface, we observe a gaussian-shaped noise distribution corresponding to typically 1-2 atoms/pixel and no pronounced structures in the images. For images close to the surface where it is possible to observe reflected, diffracted and direct light, the noise depends on the specific region of the image. In regions where there is large contrast in the resulting interference fringes, the noise can be significantly larger. When the beam angle is adjusted carefully, we obtain noise levels similar to those of images taken far from the surface. Figure \ref{fig:noiseTOF}(a) shows an absorption image of a BEC of ${~}^{87}$Rb atoms taken 5 ms after release from a trap formed 10 $\mu$m from the surface of the atom chip.  The image is taken with an imaging beam angle of incidence of $\theta \sim 1^\circ$. The modulation of the BEC density (Fig.~\ref{fig:noiseTOF}(b),(c)) is caused by inhomogeneous current flow through the atom chip wire used to form the magnetic trap \cite{Wildermuth2005b,Krueger2007}. The noise floor of the image can be estimated to be $\sim$ 2 atoms/$\mu$m rms. With an object pixel size for the image of 3.35~x~3.35~$\mu$m$^2$ we estimate a noise of $\sim$ 7 atoms/pixel row in the image of the cloud in Fig.~\ref{fig:noiseTOF}.  With a half width of $\sim$ 15~$\mu$m rms for the cloud in Fig.~\ref{fig:noiseTOF}, a pixel row corresponds to $\sim$100~$\mu$m$^2$.  The rms noise of the picture correspond to a column density $n\sim 0.07$~atoms/$\mu$m$^2$. With a shorter expansion time the detectable linear density $n_{1D}$ in this imaging setup is on the order of $n_{1d}<0.5$~atoms/$\mu$m.

\section{Detection Limits}
The information available from an absorption image is limited by the signal-to-noise ratio (SNR) which determines the minimum detectable atom number, as discussed in \cite{Wilzbach2006}.
Increasing the intensity $I_\mathrm{in}$ and imaging duration $\tau$ would lead to a decrease in the minimum detectable atom number $N_{min}$.  However, the photon recoil in the atom-light scattering process heats the atoms and they move during the imaging duration. For freely propagating atoms, this results in a limit to the imaging duration $\tau$ for the atoms to remain in a given area $A$. Increasing $I_\mathrm{in}$ to too large a value brings two problems.  Firstly, increasing $I_\mathrm{in}$ increases the heating of the atom cloud as the atom-light scattering rate is increased.  Secondly, the absorption cross section $\sigma$ can only be considered to be intensity independent for intensities that are small compared to the saturation intensity of the optical transition used for imaging: beyond this, if one wishes to extract accurate atom numbers, $I_\mathrm{in}$ must be accurately known.  In addition, CCD cameras have a quantum efficiency $\eta$ for the conversion of photons to electrons that should be taken into account when considering $N_{min}$.

As an example we estimate the limits for $^{87}$Rb and an imaging beam intensity $I_\mathrm{in}$ of $10\%$ of the saturation intensity $I_\mathrm{sat}$ of the transition (a typical value used in our experiments). For the $|F=2,m_F=2\rangle \rightarrow |F^{\prime} =3, m_F=3\rangle$ transition on the \textit{D}$_{2}$ line, ($\lambda = 780$~nm) the absorption cross section $\sigma_{lin}=0.19$~$\mu$m$^2$ ($I_\mathrm{sat (lin)} = 25$~Wm$^{-2}$) for linearly polarised light and $\sigma_{cir}=0.29$~$\mu$m$^2$ ($I_\mathrm{sat (cir)} = 17$~Wm$^{-2}$ ) for circularly polarised light \cite{Steck2001}.  Using an optical imaging system resolution of 3\,$\mu$m gives $A=\pi \times 3^{2}=28~\mu$m$^{2}$ and we obtain an approximate minimum column density $n_{min}$ (in atoms per $\mu$m$^{2}$)  as $n_{min(lin)} > 0.44~\tau^{-1/2}$ and $n_{min(cir)} > 0.35~\tau^{-1/2}$ for linear and circularly polarised light, respectively ($\tau$ (in $\mu$s) is the duration of the imaging pulse). Consequently, $N_{min(lin)} > 12~\tau^{-1/2}$ and $N_{min(cir)} > 10~\tau^{-1/2}$.

Experimentally, the imaging duration $\tau$ is usually in the region of $\sim$ 30~$\mu$s which leads to the following limits:
$n_{min(lin)} = 0.08~\mu$m$^{-2}$ and $N_{min(lin)} = 3$~atoms for linearly polarised light and $n_{min(cir)} = 0.06~\mu$m$^{-2}$ and $N_{min(cir)} = 2$~atoms for circularly polarised light, for the experimental parameters given above. These values correspond to an attenuation of the incoming imaging beam of $1.5\%$ and $1.7\%$ for linearly and circularly polarised light, respectively. To achieve $N_{min} = 1$, would require a much longer imaging pulse, by which time most of the atoms would have moved out of the area $A$ under observation.  For single atom detection by absorption, the atoms would have to be tightly confined during the imaging pulse, whereas freely propagating single atoms can be detected by fluorescence, as described in~\cite{Bucker2009}.

\section{Imaging Systems: Two-Lens Imaging Systems}\label{sec:TwoLens}
The imaging systems used for this work were composed of two lenses (Fig.~\ref{fig:translong}) located outside the vacuum chamber. By selecting the right combination of standard precision achromats we achieve diffraction-limited imaging with a resolution down to 3 $\mu$m at a working distance of 10 cm.  Higher resolution will require the first lens to be located closer to the atoms and a multi-lens design.  Examples of imaging systems implemented in our experiments are detailed below.

\textbf{Imaging System 1}.
Lens 1: Melles Griot 06LAI011 ($f=100$~mm); Lens 2: Melles Griot 01LA0339 ($f=400$~mm); 
MicroMAX:1024BFT, back-illuminated CCD with a quantum efficiency of 72$\%$ at 780~nm, from Roper Scientific. The system was designed to give a magnification of 3.9, with a NA of 0.13 and the pixel size of 13~x~13~$\mu$m$^2$ translated into 3.35~x~3.35~$\mu$m$^2$ in object space.  The imaging is diffraction-limited within a radius of $>1$~mm (in object space) from the central axis, the Airy disc had a radius of 3.7~$\mu$m.

\textbf{Imaging System 2}.
Lens 1: Melles Griot 06LAI011 ($f=100$~mm); Lens 2: Thorlabs AC508-750B ($f=750$~mm) achromat resulting in a magnification of 7.4. The NA of 0.13 gave a 3.7~$\mu$m radius of the Airy disk, the imaging was diffraction-limited within a radius of $>1$~mm in object space. The images were recorded with an Andor frame-transfer CCD camera with a pixel size of 13 x 13~$\mu$m$^2$ translating to 1.8 x 1.8~$\mu$m$^2$ in the object plane.

\textbf{Imaging System 3}.
Lens 1: Melles Griot 06LAI013 ($f=145$~mm); Lens 2: Thorlabs AC508-750B ($f=750$~mm); Princeton Instruments spectroscopy CCD camera with 1300 x 400 pixels of 20 x 20~$\mu$m$^2$. A magnification of 4.8 gave an object pixel size of 4.2~$\mu$m. With a NA of 0.09 the system was diffraction-limited within a radius of $>2$~mm from the central axis, with an Airy disk radius of 5.4~$\mu$m.

\textbf{Imaging System 4}.
Lens 1: Thorlabs AC254-150-B ($f=150$~mm); Lens 2 Thorlabs AC508-1000-B ($f=1000$~mm);  Princeton Instruments spectroscopy CCD camera with 1300 x 400 pixels of 20 x 20~$\mu$m$^2$. A magnification of 9.3 gave an object pixel size of 2.2~$\mu$m.  With a NA of 0.08 the system was diffraction-limited within a radius of $>2$~mm from the central axis, with an Airy disk size of 5.7~$\mu$m.

\section{Conclusion and Outlook}
We have discussed absorption imaging of ultracold atoms close the surface of atom chips, particularly focusing on the case where the imaging beam is purposely reflected off the atom chip surface.  We have shown that a standing wave is produced above the chip surface due to this reflection and that this must be taken into account when imaging, particularly for atom cloud sizes smaller than the standing-wave wavelength.  Having built both a simple E-field model for the light and carried out a wave-propagation simulation to demonstrate the standing wave phenomenon, we demonstrated a Fourier method that is a stable way to extract the height of the atom cloud above the atom chip surface. We also discussed other configurations for absorption imaging when using the atom chip experimental environment, before discussing experimental optimisation, detection limits and two-lens imaging systems.  Absorption imaging techniques close to the surface of an atom chips, are complimentary to other methods, including chip-based optical fibres for fluorescence detection~\cite{Heine2010}, chip-based optical cavities~\cite{Colombe2007}, and photoionization~\cite{Kraft2007}, as well as fluorescence detection far from the chip surface~\cite{Bucker2009}, each having their particular benefits and drawbacks that together make a comprehensive toolbox for atom detection in atom chip experiments.

\section{Acknowledgments}
This work was supported by the European Union (CHIMONO), the Deutsche Forschungsgemeinschaft and the Austrian FWF through the Wittgenstein prize, the CoQuS doctoral program, grant no. P20372, and the SFB FoQuS. DAS thanks the Austrian FWF Lise Meitner fellowship program. The authors thank Robert B\"ucker and Georg Winkler for interesting discussions.

% bibliography taken from .bbl file created using bibtex and osajnl style file.

\end{document}